\documentclass[twocolumn,showpacs,preprintnumbers,amsmath,amssymb]{revtex4}

\usepackage{graphicx}
\usepackage{dcolumn}
\usepackage{bm}

\begin{document}


\title{
Critical Behavior of the SDW Transition in Under-doped
Ba(Fe$_{1-x}$Co$_{x}$)$_2$As$_2$ ($x \leq 0.05$): \\ $^{75}$As NMR
Investigation }

\author{F.\ L.\ Ning$^{1}$\footnote{Electronic address: ningfl@zju.edu.cn}, M.\ Fu$^{2}$, D.\ A.\ Torchetti$^{2}$, T.\ Imai$^{2,3}$, A.\ S.\ Sefat$^{4}$, P.\ Cheng$^{5}$, B.\ Shen$^{5}$ and H.-H\ Wen$^{3,5,6}$}

\affiliation{$^{1}$Department of Physics, Zhejiang University,
Hangzhou 310027, P. R. China} \affiliation{$^{2}$Department of
Physics and Astronomy, McMaster University, Hamilton, Ontario
L8S4M1, Canada} \affiliation{$^{3}$Canadian Institute for Advanced
Research, Toronto, Ontario M5G1Z8, Canada}
\affiliation{$^{4}$Materials Science and Technology Division, Oak
Ridge National Laboratory, TN 37831, USA}
\affiliation{$^{5}$National Laboratory for Superconductivity,
Institute of Physics and Beijing National Laboratory for Condensed
Matter Physics, Chinese Academy of Sciences, Beijing 100190, P.R.
China} \affiliation{$^{6}$Center for Superconducting Physics and
Materials, National Laboratory for Solid State Microstructures,
Department of Physics, Nanjing University, Nanjing 210093, P.R.
China}

\begin{abstract}

We investigate the nature of the SDW (Spin Density Wave) transition
in the underdoped regime of an iron-based high $T_c$ superconductor
Ba(Fe$_{1-x}$Co$_{x}$)$_2$As$_2$ by $^{75}$As NMR, with primary
focus on a composition with $x=0.02$ ($T_{SDW}=99$\ K).  We
demonstrate that critical slowing down toward the three dimensional
SDW transition sets in at the tetragonal to orthorhombic structural
phase transition, $T_{s} = 105$\ K, suggesting strong interplay
between structural distortion and spin correlations.  In the
critical regime between $T_{s}$ and $T_{SDW}$, the dynamical
structure factor of electron spins $S({\bf q}, {\omega_{n}})$
measured with the longitudinal NMR relaxation rate $1/T_{1}$
exhibits a divergent behavior obeying a power-law, $1/T_{1} \propto
\Sigma_{\bf q} S({\bf q}, {\omega_{n}}) \sim
(T/T_{SDW}-1)^{-\delta}$ with the critical exponent $\delta \sim
0.33$.
\end{abstract}

\maketitle

\section{Introduction}

The discovery of superconductivity with $T_c$ as high as 28 $\sim$
55 K in iron-pnictides \cite{Kamihara, Ren, ChenSm, Rotter, Sefat}
has regenerated strong interest in the research of high temperature
superconductivity.  The parent compound of the so-called 122
ferropnictides, BaFe$_2$As$_2$, is a semi-metallic antiferromagnet;
upon cooling, BaFe$_2$As$_2$ undergoes a first-order Spin Density
Wave (SDW) transition at $T_{SDW}\sim135$\ K, accompanied by a
tetragonal to orthorhombic structural phase transition at
$T_{s}(=T_{SDW})$ \cite{Rotter, Huang, Takigawa, Iyo}.  Doping a few
\% of Co into the Fe sites of BaFe$_2$As$_2$ quickly suppresses
$T_{SDW}$\cite{Ning2, Ni}  as well as $T_{s}$\cite{Fisher}, as
summarized in Fig.\ 1.   In the lightly Co doped regime, the
structural phase transition takes place first upon cooling, followed
by the SDW transition in the orthorhombic phase\cite{Fisher,Nandi}.
Superconductivity with optimized $T_{c}\sim 25$\ K appears when
$T_{s}$ and $T_{SDW}$ are completely suppressed by 6 $\sim$ 8 \% Co
doping \cite{Sefat, Ning2, Ni, Fisher, ChenXH, Nandi,Wen}.   The
nature and origin of the SDW ordering, and its potential relation to
the superconducting mechanism, are the subject of intense debates
\cite{Johnston}.

In this work, we investigate the critical behavior of the SDW
transition and its interplay with the structural transition in
lightly Co doped single crystals of Ba(Fe$_{1-x}$Co$_{x}$)$_2$As$_2$
with $x=0.02$, 0.04, and 0.05 based on  $^{75}$As NMR measurements.
We will place our primary focus on a composition with $x=0.02$;
thanks to its relatively sharp NMR lines, experimental
characterizations of structural and SDW phase transitions are
straightforward for this composition.  We demonstrate
that the structural transition at $T_{s}=105$~K triggers the
critical slowing down of spin dynamics toward the three dimensional
SDW transition at $T_{SDW}=99$~K.  We found that the critical
exponent for the divergence of the dynamical structure factor of
electron spins, $S({\bf q}, {\omega_{n}})$, near the SDW transition
is different from $\delta = 1/2$ often attributed to itinerant
electron magnetism, such as metallic Cr \cite{Masuda}. Instead, we
found $\delta \sim 0.33$.  This value is nearly identical with the
case of a Mott insulator CuO with $\delta = 0.33\pm0.01$
\cite{Itoh}, and is in reasonable agreement with the theoretically
predicted value of $\delta = \nu/2 \sim 0.35$ for insulating
three-dimensional (3D) Heisenberg antiferromagnets \cite{Halperin,
Kawasaki, Hohenemser, Lovesey}.  Here $\nu\sim0.7$ is the critical
exponent for the spin-spin correlation length $\xi$, and $\xi \sim
(T/T_{SDW}-1)^{-\nu}$.  We also demonstrate that Co doping enhances
the density of states $D(E_F)$ of the reconstructed Fermi surfaces
below $T_{SDW}$ roughly in proportion to $x$, based on the
enhancement of $1/T_{1}T$ at low temperatures.

The rest of this paper is organized as follows.  In section II, we
will briefly describe experimental procedures.  In section III, we
will discuss our results in the paramagnetic state above $T_{SDW}$,
followed by brief discussions about the SDW ordered state.  We will
conclude in section IV.

\begin{figure}[t]
\includegraphics[width=2.8in]{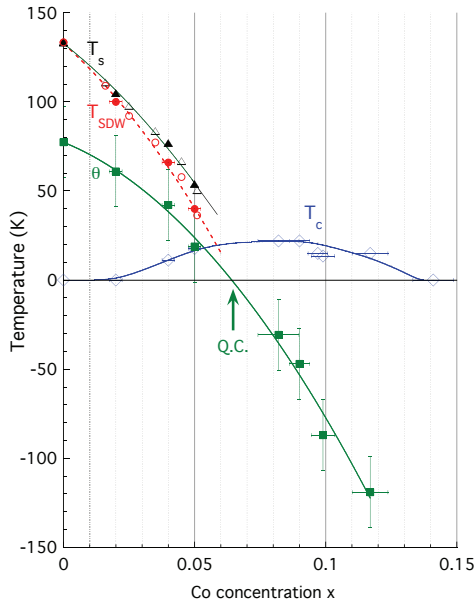}\\
\caption{\label{Fig.1} (Color Online) The superconducting transition
temperature $T_c$ of the present series of samples ($\diamondsuit$)
\cite{Ning2, Ning3}; the tetragonal to orthorhombic structural phase
transition temperature $T_{s}$ determined from the onset of the NMR
line broadening ($\blacktriangle$); the SDW transition temperature
$T_{SDW}$ determined from the power-law fit of $1/T_{1}$ in the
critical region ($\bullet$); and Weiss temperature $\theta$ of the
imaginary part of the staggered spin susceptibility $\chi"({\bf q},
\omega_{n})$ determined by the mean-field fit of the $1/T_{1}T$ in
the tetragonal phase ($\blacksquare$) \cite{Ning3}.   For
comparison, we also show $T_{s}$ ($\triangle$) and $T_{SDW}$
($\circ$) as determined by anomalies observed in resistivity
\cite{Fisher}.  Upward arrow marks the magnetic quantum critical
point, $x_{c}\sim 0.065$ \cite{Ning3}.  All solid lines are guides
for the eye. }
\end{figure}
\section{EXPERIMENTAL METHODS}

We grew single crystals of Ba(Fe$_{1-x}$Co$_{x}$)$_2$As$_2$ from
FeAs flux \cite{Sefat, Wen}.  We carried out NMR measurements using
the standard pulsed NMR techniques.  For $x=0.02$, we cleaved a
small piece of shiny crystal from a much larger boule used for our
previous report \cite{Ning4}.  The total mass of the smaller crystal
used for the present work is about $\sim 7$\ mg.  It was necessary
to use the smaller piece to ensure high homogeneity of the sample.
In fact, we found no evidence for a stretched recovery of
$T_{1}$ \cite{Dioguardi} in our small homogeneous crystal of
$x=0.02$, contrary to an earlier report that a $x=0.022$ crystal
\cite{Curro} and lightly doped LaFeAsO$_{1-x}$F$_{x}$ crystals
\cite{Hammerath} exhibit a large distribution of $T_{1}$, which
implies a large distribution of $T_{SDW}$.  From the sharpness of
the divergent behavior of $1/T_{1}$ and the NMR linewidth, we
estimate the upper bound of the distribution of $T_{s}$ and
$T_{SDW}$ as little as $\sim 0.5$\ K in our small $x=0.02$ crystal.
Moreover, we could resolve the fine structures of the NMR lineshapes
in the magnetically ordered state below $T_{SDW}$ (see Fig.~2(b) and
(c) below), which we were unable to detect in our earlier study
using a larger, inhomogeneous crystal \cite{Ning4}.  Due to the poor
signal to noise ratio arising from the small volume of the crystal
and long relaxation time $T_1$, the NMR data acquisition is
extremely time consuming below $T_{SDW}$; it took up to $\sim$10
days of continuous signal averaging to complete one set of NMR
lineshape measurements at a given temperature.

Small single crystal samples used for other compositions with $x=
0.04$ and $x = 0.05$ are identical with those used in our previous
studies \cite{Ning2, Ning3}.  We found stretched forms of $T_{1}$
recovery only for $x = 0.05$ below $\sim 70$~K, analogous to the
earlier report \cite{Curro}.  It is worth recalling that Co
substitution is known to suppress spin fluctuations {\it locally} at
Co sites, as evidenced by temperature independent $1/T_{1}T$
observed at Co sites at low temperatures \cite{Ning1}.  A level of
distribution in the electronic properties in the alloyed samples of
Ba(Fe$_{1-x}$Co$_{x}$)$_2$As$_2$ is therefore naturally expected, as
we demonstrated earlier from the variation of $1/T_{1}$ within a
single NMR peak of a given composition \cite{Ning2}.   But none of
the key findings and conclusions in the present work rely on the
$x=0.05$ sample at low temperatures, and hence the issue of the
inhomogeneity induced by Co substitution is beyond the scope of the
present work.
\section{Results and Discussions}
\subsection{$^{75}$As NMR lineshape, width and Knight shift}

In Fig.\ 2(a), we present a representative field-swept $^{75}$As NMR
lineshape  of Ba(Fe$_{0.98}$Co$_{0.02}$)$_2$As$_2$ observed at a
fixed NMR frequency of $\omega_{n}/2\pi = 43.503$\ MHz in the
paramagnetic state above $T_{SDW}$.  In general, the nuclear spin
Hamiltonian can be expressed as a summation of the Zeeman and
nuclear quadrupole interaction terms,
\begin{equation}
\label{1} \ \ H =-\gamma_n \textit{h} \vec{B} \cdot \vec{I}+
\frac{\textit{h} \nu_Q^c}{6}\{3I_z^2- I(I+1) +
\frac{1}{2}\eta(I_+^2+I_-^2)\},
\end{equation}
where the $^{75}$As nuclear gyromagnetic ratio is $\gamma_{n}/2\pi =
7.2919$\ MHz/T, $\textit{h}$ is Planck's constant, and $\vec{I}$
represents the nuclear spin.  Since $^{75}$As has nuclear spin
$I=3/2$, we observe three transitions from $I_{z} = \frac{2m-1}{2}$
to $\frac{2m+1}{2}$ (with $m=-1$, 0, and $+1$) in the NMR lineshape:
the sharp central peak arises from the $I_{z}=-1/2$ to +1/2
transition ($m=0$); additional two broad satellite peaks arise from
$I_{z}=\pm3/2$ to $\pm1/2$ transitions ($m=\pm1$), separated by
$^{75}\nu_Q^{c}$.  The nuclear quadrupole interaction frequency
$\nu_Q^c$ along the $c$-axis is proportional to the Electric Field
Gradient (EFG) at the observed $^{75}$As site, and $\eta$ is the
asymmetry parameter of the EFG, $\eta$ =
$|\nu_Q^a-\nu_Q^b|/\nu_Q^c$.  Due to the tetragonal symmetry at the
$^{75}$As sites, $\eta=0$ above $T_{s}$.   Co doping induces
substantial disorder in the lattice, reflected on the distribution
of $^{75}\nu_Q^{c}$.

$\vec{B}$ is the summation of the external field $\vec{B}_{ext}$ and
the time averaged hyperfine fields from nearby electron spins,
$\vec{B}_{hf}$, i.e. $\vec{B} = \vec{B}_{ext} + \vec{B}_{hf}$.  In
the paramagnetic state, the central peak frequency is slightly
shifted (i.e. ``Knight shift'') due to small hyperfine fields
induced by polarized electron spins nearby.  Since the spin
polarization induced by $\vec{B}_{ext}$ is proportional to spin
susceptibility $\chi_{spin}$, we can measure the latter by
accurately determining the central peak position \cite{Jaccarino}.
In the SDW ordered state, static $\vec{B}_{hf}$ induced by ordered
magnetic moments in the vicinity of the observed $^{75}$As nuclear
spins dramatically affect the NMR lineshapes, as shown in Fig.2\ (b)
and (c).   We will come back to this point below in section C.

\begin{figure}[t]
\includegraphics[width=3in]{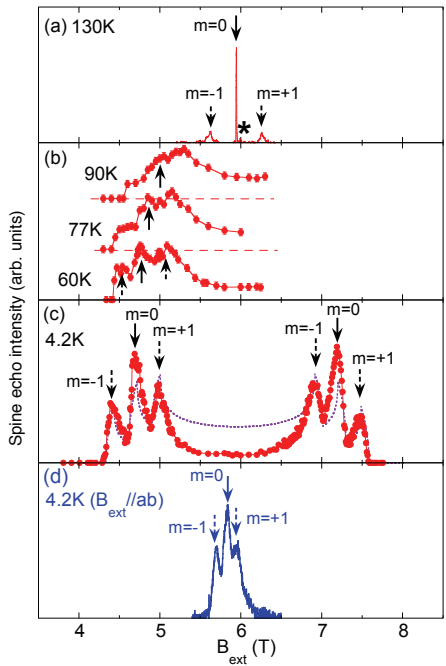}\\
\caption{\label{Fig.2} (Color Online) (a)-(c) Field-swept NMR
lineshapes of Ba(Fe$_{0.98}$Co$_{0.02}$)$_2$As$_2$ obtained for a
fixed frequency of 43.503\ MHz under the external magnetic $B_{ext}$
applied along the c-axis.  Solid and dashed arrows represent the
central ($m=0$) and satellite ($m=\pm1$) transitions. $\star$ in (a)
marks a weak $m=0$ peak arising from $^{75}$As(1) sites located at
the nearest neighbor of Co sites (see \cite{Ning3} for details).
The vertical origin for the data points at 77\ K and 90\ K is
shifted for clarity.  The dotted curve in (c) represents an attempt
to fit the lineshape at 4.2\ K with a sinusoidally modulated static
hyperfine fields expected for an incommensurate SDW.  (d) The
lineshape obtained for $B_{ext} || ab$. }
\end{figure}

We summarize the temperature dependence of the paramagnetic NMR
Knight shift $^{75}K$ and the FWHM (Full Width at Half Maximum) of
the central peak frequency in Fig.\ 3 and 4, respectively.  To
ensure high accuracy, we conducted these measurements by taking the
FFT of the spin echo envelope in a fixed magnetic field.   The NMR
Knight shift, $^{75}K = A_{hf}\chi_{spin} + K_{chem}$, probes the
local spin susceptibility $\chi_{spin}$ via hyperfine coupling
$A_{hf}$; $K_{chem}$ ($\sim 0.2$~\% or less for $x=0.02$)  is a
temperature independent chemical shift \cite{Ning1}.  Our new
results of  $^{75}K$ in Ba(Fe$_{0.98}$Co$_{0.02}$)$_2$As$_2$ are
analogous to those observed for other compositions \cite{Ning1,
Ning2, Ning3}: $^{75}K$ decreases with temperature, and tends to
level off near $\sim 100$\ K \cite{Oh}.  See \cite{Ning3} for
detailed analysis of $^{75}K$ based on fitting the data with a
pseudo gap $\Delta_{PG}/k_{B}\sim 450$\ K.

One interesting aspect of Fig.\ 3 is that $^{75}K$  exhibits a
noticeable drop below $105.0\pm0.5$\ K for
Ba(Fe$_{0.98}$Co$_{0.02}$)$_2$As$_2$.  This anomaly is accompanied
by a sudden onset of the divergent behavior of FWHM, as shown in
Fig.\ 3.  We note that FWHM indeed diverges below
$T_{SDW}=99.0\pm0.5$\ K, where the emergence of static hyperfine
magnetic field $B_{hf}$ splits the NMR line in the SDW ordered
state, as shown in Fig.\ 2(b) and (c).
We found analogous anomalies of  $^{75}K$ and FWHM for Co 4\% and
5\% doped samples at $77\pm2$\ K and $55\pm2$\ K, respectively, as shown in Figs.\ 3 and 4.  We
summarize the concentration dependence of these anomalies in Fig.\
1.  Clearly, these anomalies are related to the structural phase
transition at $T_{s}$ \cite{Fisher, Nandi}.
\begin{figure}[t]
\includegraphics[width=3in]{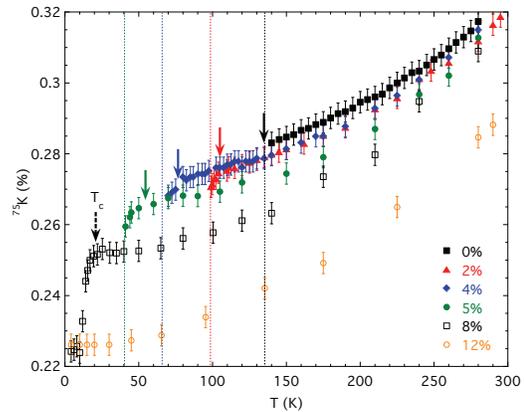}\\
\caption{\label{Fig.3} (Color Online) The $^{75}$As NMR Knight shift
$^{75}K$ observed in $B_{ext} || c$ for
Ba(Fe$_{0.98}$Co$_{0.02}$)$_2$As$_2$ is compared with the case of
$x=0$, 0.04, 0.05, 0.08, and 0.12 \cite{Ning1,Ning2,Ning3}.
Downward solid arrows mark $T_{s}$, while vertical dotted line
represents $T_{SDW}$ as determined by the divergence of $1/T_{1}$.
Notice the downturn of $^{75}K$ below $T_{s}$ for $x=0.02$, 0.04,
and 0.05. }
\end{figure}
\begin{figure}[t]
\includegraphics[width=3in]{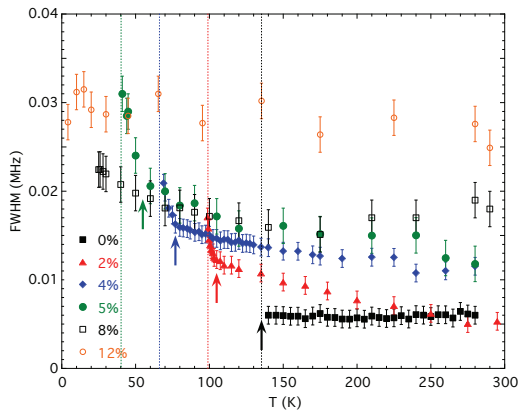}\\
\caption{\label{Fig.4} (Color Online) FWHM of the $^{75}$As NMR
central transition in an external magnetic field of
$B_{ext}\simeq8.3$\ T applied along the crystal c-axis.  Upward
solid arrows mark $T_{s}$, while vertical dotted line represents
$T_{SDW}$.  Notice the sudden upturn of FWHM below $T_{s}$ for
$x=0.02$, 0.04, and 0.05.  The plot of FWHM data is discontinued at
$T_{SDW}$ because NMR lines split below $T_{SDW}$, as shown in Fig.\
2(b) and (c). }
\end{figure}

Having identified the signature of the structural phase transition
at $T_{s}$ in our NMR data for the central transition, we also
searched for an anomaly in the nuclear quadrupole frequency
$^{75}\nu_{Q}^{c}$ by measuring the splitting between the central
and satellite peaks.  We recall that, in typical second order
structural phase transitions such as the high temperature tetragonal
to low temperature orthorhombic phase transition in the undoped and
Sr-doped La$_{2}$CuO$_{4}$ high $T_c$ cuprates, one could even
observe a $\lambda$-like kink in the temperature
dependence of $\nu_{Q}^{c}$ \cite{Imai1993}.  We summarize our
results for Ba(Fe$_{0.98}$Co$_{0.02}$)$_2$As$_2$ in Fig.\ 5. (High
precision determination of $^{75}\nu_{Q}^{c}$ is rather difficult
for higher Co concentrations, because the satellite peaks become
very broad due to disorder \cite{Ning1}.)  In the case of undoped
BaFe$_2$As$_2$, $^{75}\nu_{Q}^{c}$ exhibits a step at the first
order structural transition $T_{s}=135$\ K \cite{Takigawa}, but we
find practically no anomaly at $T_{s}=105.0$\ K for the Co 2\%
sample.  In general, when the lattice contracts with decreasing
temperature, the lattice contribution to the electric field gradient
(EFG), and hence to $\nu_{Q}$, increases.  Our finding that
$^{75}\nu_{Q}^{c}$ smoothly decreases with temperature might be an
indication that there is a sizable on-site ionic contribution with
an opposite sign.

It is not clear why $^{75}\nu_{Q}^{c}$ does not exhibit a clear
anomaly at $T_{s}$ for the Co 2\% doped sample.  One possible
scenario is that the influence of structural distortion on $^{75}$As
sites becomes so subtle  under the presence of Co dopants that the
change of $^{75}\nu_{Q}^{c}$ also becomes extremely small.  We also
recall that softening of the lattice stiffness begins at unusually
high temperatures in  Ba(Fe$_{1-x}$Co$_{x}$)$_2$As$_2$, and has been
speculated to be the consequence of antiferromagnetic correlations
\cite{Mandrus, Uchida, FernandesPRL}.  Perhaps the effects of
orthorhombic distortion on $^{75}\nu_{Q}^{c}$ appear progressively
from much higher temperature than $T_{s}$.  In any event, the
absence of a strong signature of  structural anomaly in the
temperature dependence of  $^{75}\nu_{Q}^{c}$ at $T_{s}$ excludes
the possibility that anomalies observed below $T_{s}$ in Figs.\ 3
and 4 are a consequence of the subtle changes in the second order
quadrupole effects.  In fact, we confirmed that the FWHM is
approximately proportional to the magnitude of the applied magnetic
field, hence the divergent behavior of FWHM below $T_{s}$ is the
consequence of magnetic effects.  We recall that the NMR line
broadened by the second order quadrupole effects would be inversely
proportional to the magnetic field instead.

Quite generally, divergence of the NMR linewidth precedes a magnetic
phase transition through the divergence of dynamical spin
susceptibility in the critical regime \cite{Jaccarino}.   We also
recall that the NMR Knight shift $^{75}K$ reflects local
paramagnetic spin susceptibility $\chi_{spin}$, hence the downturn
in the temperature dependence of $^{75}K$ below $T_{s}$ is also
consistent with suppression of $\chi_{spin}$ due to strong
antiferromagnetic short-range order.  Thus our findings in both
Fig.\ 3 and 4 suggest that the structural phase transition at
$T_{s}$ drives the onset of strong 3D antiferromagnetic short range
order.  This point is more vividly demonstrated through the
divergent behavior of $1/T_{1}$ in the next section.

\begin{figure}
\includegraphics[width=3in]{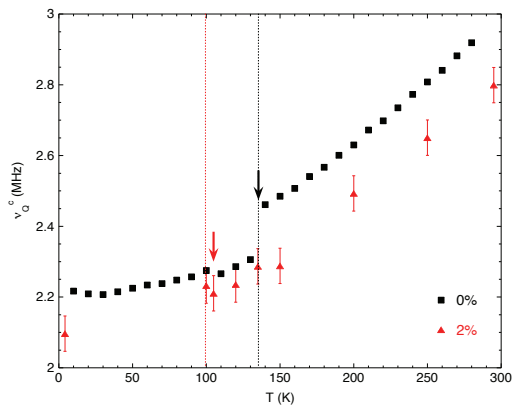}\\
\caption{\label{Fig.5} (Color Online) The c-axis component of the
$^{75}$As nuclear quadrupole frequency $^{75}\nu_{Q}^{c}$ in
($\blacktriangle$) Ba(Fe$_{0.98}$Co$_{0.02}$)$_2$As$_2$ (present
work), and ($\blacksquare$) undoped BaFe$_2$As$_2$ \cite{Takigawa}.  Downward arrows mark $T_{s}$, while vertical dotted line represents $T_{SDW}$.
We were unable to determine $^{75}\nu_{Q}^{c}$  accurately below
$T_{SDW}=99.0$\ K except at 4.2\ K due to extremely broad line
profiles (see Fig.\ 2(b)). }
\end{figure}

\subsection{Critical spin dynamics near $T_{SDW}$}

In Fig.\ 6, we present $^{75}$As nuclear spin-lattice relaxation
rate $1/T_{1}$ divided by temperature $T$, i.e.  $1/T_{1}T$,
observed for Ba(Fe$_{0.98}$Co$_{0.02}$)$_2$As$_2$.  $1/T_{1}T$
measures wave-vector {\bf q} integral of the imaginary part of the
dynamical electron spin susceptibility $\chi"({\bf q}, \omega_{n})$
weighted by the hyperfine form factor $|A_{hf}({\bf q})|^{2}$
\cite{Shannon}.  In the case of undoped  BaFe$_2$As$_2$, $1/T_{1}T$
does not show divergent behavior at $T_{SDW}$ expected for second
order magnetic phase transitions; instead,  $1/T_{1}T$ shows a step
at 135\ K because the SDW transition is first order \cite{Takigawa}.
In contrast, $1/T_{1}T$ observed for Co 2\% doped sample exhibits
strongly divergent behavior near $T_{SDW}$ in the geometry of
$B_{ext} || ab$.  In this configuration, $1/T_{1}T$ probes
fluctuations of hyperfine fields both along the c-axis and ab-plane.
The divergent signature is less prominent for $B_{ext} || c$,
because $1/T_{1}T$ probes fluctuating hyperfine fields only within
the ab-plane, and the transferred hyperfine field $A_{hf}({\bf q})$
becomes vanishingly small for staggered wave vectors in this
configuration \cite{Takigawa, Johnston, Shannon}.  In other words,
it is advantageous to use the $B_{ext}~||~ab$ geometry to probe the
critical behavior of the SDW transition.
\begin{figure}[t]
\includegraphics[width=3in]{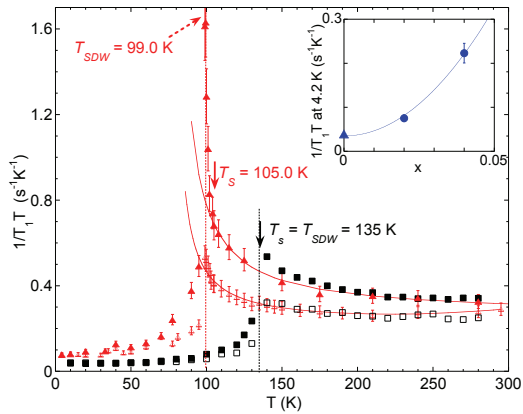}\\
\caption{\label{Fig.6} (Color Online) $1/T_{1}T$ observed for
Ba(Fe$_{0.98}$Co$_{0.02}$)$_2$As$_2$ under the external magnetic
field $B_{ext}~||~ab$ ($\blacktriangle$) or $B_{ext}~||~c$
($\triangle$).  For comparison we also show the results of
BaFe$_2$As$_2$ for $B_{ext}~||~ab$ ($\blacksquare$) and
$B_{ext}~||~c$ ($\square$) \cite{Takigawa}.   Vertical dotted lines
represent $T_{SDW}$, while solid arrows mark $T_{s}$.  Solid curves
are a Curie-Weiss fit (see main text).  Notice that the Curie-Weiss
fit breaks down at $T_{s}=105.0$\ K, and $1/T_{1}T$ begins to blow
up toward $T_{SDW}=99.0$\ K.    Inset: the concentration $x$
dependence of $1/T_{1}T$ at 4.2\ K for $B_{ext}~||~c$.  The solid
curve is a parabolic fit. }
\end{figure}
\begin{figure}[t]
\includegraphics[width=3in]{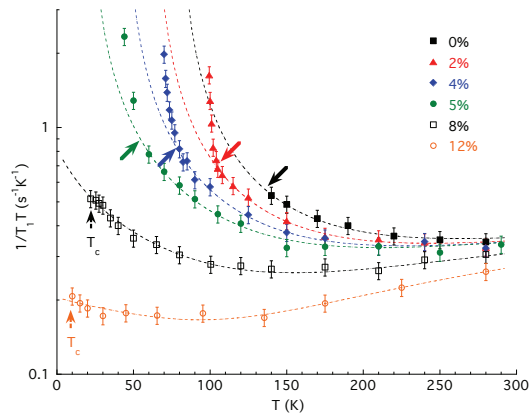}\\
\caption{\label{Fig.7} (Color Online) A semi-log plot of $1/T_{1}T$
observed for Ba(Fe$_{1-x}$Co$_{x}$)$_2$As$_2$ with $B_{ext}||ab$.
For clarity, we show data points only above $T_{SDW}$ for $x\leq
0.05$, and above $T_{c}$ for $x=0.08$ and 0.12.  Dashed curves
represent a phenomenological Curie-Weiss fit, incorporating a
background term due to a pseudo-gap (see main text).  Slanted solid
arrows mark $T_{s}$ for $x=0 \sim 5$\%, while vertical dashed arrows
show $T_{c}$ for $x=8$\% and 12\%. }
\end{figure}

Accordingly, in what follows, we focus our attention on $1/T_{1}T$
measured in $B_{ext} || ab$.  In Fig.\ 7, we show $1/T_{1}T$ in a
semi-log scale for various Co concentrations.  To avoid confusion,
we show only the results above $T_{SDW}$ (for $x \leq 5$\%) or
$T_{c}$ (for $x=8$\% and 12\%).  Also presented is a
phenomenological Curie-Weiss fit to an empirical equation, $1/T_{1}T
= C/(T-\theta) + A \cdot exp(-\Delta_{PG}/k_{B}T)$ \cite{Ning3}. $C$
and $A$ are fitting parameters, and $\theta$ is the Weiss
temperature of the staggered spin susceptibility $\chi"({\bf q},
\omega_{n})$ near the ordering vector.  The concentration dependence
of $\theta$ thus obtained is summarized in Fig.\ 1.  Note that we
have reversed the sign convention for $\theta$ in the present work
(i.e. $-\theta$ in Fig.\ 1 corresponds to $+\theta$ in
\cite{Ning3}).   The second, activation term in the fit represents
the background contributions which decrease with temperature,
reflecting the pseudo-gap like signature commonly observed for
iron-pnictide and iron-selenide superconductors \cite{Ahilan, Nakai,
Ning3, Imai, DaveKFeSe, Zheng}. As already discussed in detail in
\cite{Ning3}, the phenomenological Curie-Weiss fit captures the
temperature and concentration dependence of  $\chi"({\bf q},
\omega_{n})$ remarkably well, including the new results for the Co
2\% doped sample.  The Curie-Weiss behavior of $1/T_{1}T$ reflects
the fact that, upon cooling, short-range antiferromagnetic
correlations slowly grow toward $T_{SDW}$.  $\theta$ reverses its
sign above the quantum critical point $x_{c} \sim 0.065$, which
implies that Fe spins are not destined to order above $x_{c}$.
Remarkably, the optimally superconducting composition with the
maximum $T_{c}\sim 25$\ K is located in the vicinity of $x_{c}$,
hinting at the link between the superconducting mechanism and spin
fluctuations \cite{Ning3}.

Another important feature of Figs.\ 6 and 7 which we did not discuss
explicitly in \cite{Ning3} is that the phenomenological Curie-Weiss fit breaks
down below $T_{s}$.  Extrapolation of the fit to below $T_{s}$
underestimates the data points near the SDW phase transition for Co
2\%, 4\%, and 5\%, and strong divergent behavior sets in at $T_{s}$.
In other words, the three dimensional short range order sets in at
the tetragonal to orthorhombic structural phase transition, which is
prerequisite to the critical slowing down of spin fluctuations
toward the eventual three dimensional SDW order. Analogous interplay
between the spin and lattice degrees of freedom was also observed
for LaFeAsO \cite{Fu}.

In Fig.\ 8, we plot $1/T_{1}$ of three underdoped compositions on a
linear scale.  We note that $1/T_{1} \propto \Sigma_{\bf q} |
A_{hf}({\bf q}) |^{2} S({\bf q}, {\omega_{n}})$, where $S({\bf q},
{\omega_{n}})$ is the dynamical structure factor.  $1/T_{1}$ is a
very convenient probe to study the critical dynamics of $S({\bf q},
{\omega_{n}})$ in the immediate vicinity of magnetic phase
transitions, because (i) one can probe the dynamics at extremely low
energy ($\hbar\omega_{n} \sim \mu eV$), and (ii) the wave-vector
integral is automatically done.  Below $T_{s}$, we can fit the
critical dynamics with a power-law, $1/T_{1} \propto
(T/T_{SDW}-1)^{-\delta}$.  We determined $T_{SDW}$ and the critical
exponent $\delta$ based on the best fit.    The resultant values of
$T_{SDW} = 99.0$\ K (Co 2 \%), 68.9\ K (Co 4 \%), and 42.3\ K (Co
5\%) are summarized in Fig.\ 1.  The best fit also resulted in the
critical exponent $\delta = 0.329$ for Co 2 \%, and 0.317 for Co
4\%. The aforementioned distribution of $1/T_{1}$ below $\sim 70$~K
for Co 5\% makes it difficult to estimate $\delta$ with high
accuracy, but the observed temperature dependence is consistent with
$\delta \simeq 0.33$.  The inset of Fig.\ 8 shows a log-log plot of
$1/T_{1}$ as a function of the reduced temperature $(T/T_{SDW}-1)$.
The common slope in the vicinity of the SDW transition indicates
that the SDW transition of all three compositions belong to the same
universality class, and the critical exponent is given by $\delta
\sim 0.33$.  This value is close to $\delta = 0.33\pm0.01$ observed
for a Mott-insulator CuO \cite{Itoh} in the vicinity of the N\'eel
transition at $T_{N}=229$\ K, and consistent with the prediction for
insulating three dimensional Heisenberg antiferromagnets, $\delta
\sim 0.35$ \cite{Halperin, Hohenemser, Kawasaki, Lovesey}.
\begin{figure}[t]
\includegraphics[width=3in]{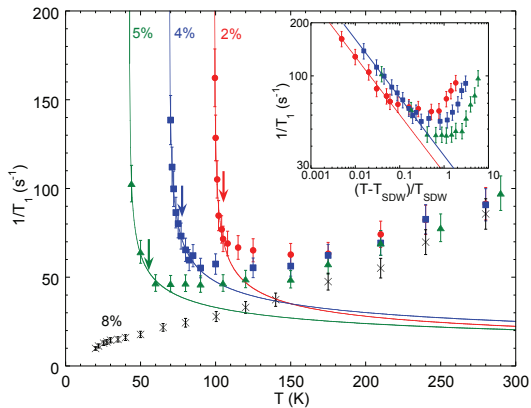}\\
\caption{\label{Fig.8} (Color Online) Power-law fits of $1/T_{1}$ in
the critical region of underdoped Ba(Fe$_{1-x}$Co$_{x}$)$_2$As$_2$.
Dotted curve and solid arrows mark $T_{s}$ and $T_{SDW}$,
respectively.  Inset: Log-log plot of $1/T_{1}$ as a function of the
reduced temperature, $(T/T_{SDW}-1)$.  Solid lines represent a power
law behavior in the critical region with $\delta=0.33$. }
\end{figure}

\subsection{Ordered moments}
\begin{figure}[tb]
\includegraphics[width=3in]{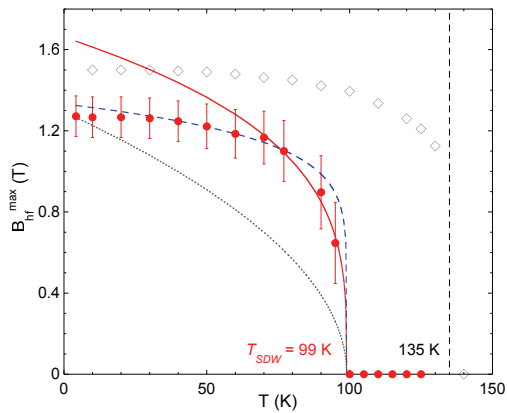}\\
\caption{\label{Fig.9} (Color Online) The temperature dependence of
$B_{hf}^{max}$ for Ba(Fe$_{0.98}$Co$_{0.02}$)$_2$As$_2$($\bullet$).
For comparison, we also show $B_{hf}^{c}$ reported  for
BaFe$_2$As$_2$ ($\diamond$) in \cite{Takigawa}. The solid line shows
$B_{hf}\sim(99.0 - T)^\beta$ with a critical exponent $\beta$ =
0.30. The dotted and dashed lines are for $\beta = 0.5$ and 0.125
plotted by fixing at two ends (T = 0 and 99.0\ K), respectively. }
\end{figure}

In Fig.\ 2(b) and (c), we show the effects of SDW ordering on the
field swept NMR lineshapes  of Ba(Fe$_{0.98}$Co$_{0.02}$)$_2$As$_2$
with $B_{ext} || c$.  We confirmed the symmetrical nature of the
lineshape at 4.2\ K, as expected, hence only the lower field half of
the lineshapes was measured in the intermediate temperature range
between 4.2\ K and $T_{SDW}$.  Below $T_{SDW}$, the entire $^{75}$As
NMR lineshape begins to split.  As noted first by Kitagawa et al. in
the case of undoped BaFe$_2$As$_2$\cite{Takigawa}, this is because
the static hyperfine magnetic field $B_{hf}$ at $^{75}$As sites
arising from the ordered Fe moments within the Fe layers point
toward the +c or -c axis.  For this reason, the overall NMR
lineshape shifts only slightly without exhibiting a splitting under
the configuration of $B_{ext} || ab$, as shown in Fig.\ 2(d).

While the observed NMR lineshapes below $T_{SDW}$ bear similarities
with the case of undoped BaFe$_2$As$_2$, there is one major
difference \cite{Ning4}: our NMR lineshapes in Fig.\ 2(b) and (c)
exhibit a continuum in the middle.  The integrated intensity between
$B_{ext}$ = 5.474 to 6.475 T accounts for $\sim$8.5 \% of the
overall intensity. This implies that $\sim$8.5 \% of $^{75}$As
nuclear spins experience $|B_{hf}| \leq 0.5$\ T, while the maximum
value of the hyperfine field reaches $B_{hf}^{max} = 1.27$\ T at
4.2\ K.  Our attempt to fit the observed lineshape with one
dimensional incommensurate modulation $B_{hf} = B_{hf}^{max} \cdot
sin (\vec{q}\cdot \vec{x})$, where $\vec{q}$ represents the
incommensurate SDW ordering vector, is unsatisfactory, as shown in
Fig.\ 2(c).   Notice that the calculated results grossly
overestimate the spectral weight in the middle part of the
lineshape.  In view of the fact that the integrated intensity of the
$^{75}$As(1) sites with a Co atom in one of their four nearest
neighbor Fe sites also accounts for approximately $\sim$7.5 \% of
the intensity (see $\ast$ in Fig.~2(a)) \cite{Ning3}, the continuum
in the middle part of the NMR lineshape may arise primarily from
$^{75}$As(1) sites.  That is, Co dopants may be suppressing the Fe
magnetic moments locally.  It has been shown by neutron scattering
that the SDW is commensurate with the lattice up to x = 0.056
\cite{Pratt,Pratt2}. Base on our NMR data, we can not prove or
disprove the incommensurability at x = 0.02. We note that similar
$^{75}$As lineshapes have been observed in the lightly doped regime
of Ba(Fe$_{1-x}$Ni$_{x}$)$_2$As$_2$ ($x$ = 0.0072 and 0.016)
\cite{Curro}.

We summarize the temperature dependence of $B_{hf}^{max}$ in Fig.\
9.  $B_{hf}^{max}$ remains approximately constant up  to $\sim 30$\
K, then decreases continuously toward $T_{SDW}= 99.0$\ K.  This
behavior is markedly different from the first order commensurate SDW
transition in BaFe$_2$As$_2$ \cite{Takigawa}; $B_{hf}$ decreases
discontinuously at $T_{SDW}$ = 135 K in the latter.  By fitting the
temperature dependence of $B_{hf}^{max}$ between 70 K
($\sim0.7~T_{SDW}$) and $T_{SDW}= 99.0$\ K to a power law,
$B_{hf}^{max}$ $\sim$ $(T_{SDW}-T)^\beta$ with a fixed $T_{SDW}$ =
99.0\ K, we obtain the critical exponent $\beta \sim 0.3$.  Very
broad lineshapes make accurate determination of $B_{hf}^{max}$
difficult near $T_{SDW}$, hence we were unable to eliminate the
large uncertainties of $\beta$.  Nonetheless, it is worth pointing
out that  $\beta \sim 0.3$ is consistent with the expectation from
the Heisenberg model, $\beta=0.37$, but different from the
mean-field value, $\beta=0.5$.

Turning our attention to the magnitude of the ordered moment
$\mu_{eff}$ at 4.2\ K as a function of $x$, we compare NMR results
with those obtained from neutron scattering in Fig.\ 10.  Since
$B_{hf}$ has a distribution under the presence of Co dopants, we
plot both the maximum value and the center of gravity of the
hyperfine field, $B_{hf}^{max}$ and $B_{hf}^{C.G.}$, respectively,
in Fig.\ 10(b).  We recall that $\mu_{eff} = 0.87\mu_B$  at 4.2\ K
for the parent compound BaFe$_2$As$_2$ \cite{Huang}, and Co doping
suppresses $\mu_{eff}$ \cite{Pratt, Lester}, as summarized in
Fig.~10(a).  On the other hand, $B_{hf}^{max} = B_{hf}^{C.G.} =
1.5$\ T observed earlier for BaFe$_2$As$_2$ \cite{Takigawa} is
gradually suppressed by Co doping.  Our results of $B_{hf}$ smoothly
extrapolate to the critical concentration as determined from
the analysis of $1/T_{1}T$ in Fig.~1, $x_{c}\sim 6.5$\%.
\begin{figure}[tb]
\includegraphics[width=3.2in]{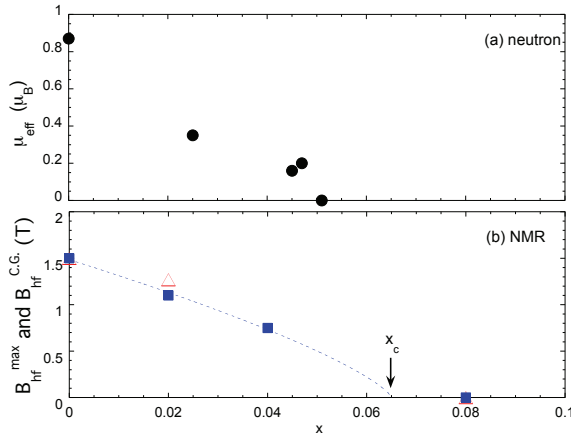}\\
\caption{\label{Fig.10} (Color Online) (a) ($\bullet$): The
concentration $x$ dependence of the magnitude of the ordered moment
$\mu_{eff}$ in for Ba(Fe$_{1-x}$Co$_{x}$)$_2$As$_2$ as determined by
neutron diffraction measurements \cite{Huang, Pratt, Lester}.  (b)
The maximum $B_{hf}^{max}$ ($\triangle$) and center of gravity
$B_{hf}^{CG}$ ($\blacksquare$) of the distribution of the hyperfine
field.  The NMR result for $x$ = 0 is from \cite{Takigawa}, while
$x_{c} \sim 0.065$ was determined from the Curie-Weiss fit of the
$1/T_{1}T$ data \cite{Ning3} (see Fig.~1).  The dotted curve is a
guide for eyes. }
\end{figure}
\subsection{Low energy spin excitations below $T_{SDW}$}

In Fig.\ 6, we show the temperature dependence of $1/T_{1}T$ below
$T_{SDW}$.  Our results show a typical $\lambda$-like temperature
dependence in the vicinity of the SDW transition.  In insulating
antiferromagnets, the low temperature behavior of $1/T_{1}T$ is
usually dominated by multi magnon Raman processes, and $1/T_{1}T$
decreases very quickly \cite{Jaccarino}.  In the present case,
however,  as we approach the base temperature of 4.2\ K, $1/T_{1}T$
levels off to a constant value of $1/T_{1}T\sim 0.08$ (s$^{-1}
K^{-1}$).  Analogous behavior was previously reported also for the
undoped parent phase BaFe$_2$As$_2$, and was attributed to the
Korringa process arising from low energy electron-hole pair
excitations at the reconstructed Fermi surface \cite{Takigawa}.  We
summarize the values of $1/T_{1}T$ observed at 4.2\ K as a function
of the doping content $x$ in the inset of Fig.\ 6,  including our
preliminary results for $x = 0.04$ \cite{Ning4}.  Interestingly,
three data points fit nicely with a parabolic function of $x$.  If
the sizable magnitude of $1/T_{1}T$ at 4.2\ K indeed arises from the
Korringa process, we expect $1/T_{1}T \propto D(E_F)^2$, where
$D(E_F)$ is the density of states.  That is, the observed parabolic
increase of $1/T_{1}T$ implies that $D(E_F)$ increases roughly
linearly with $x$.  We note that if we apply a simple rigid band
picture to the reconstructed Fermi surfaces, simple dimensional
analysis of $E_{F}$ and $D(E_{F})$ in three dimensions would lead to
$D(E_{F})\propto x^{1/3}$ instead, where $x$ is the number of
conduction electrons.

\section{SUMMARY AND CONCLUSIONS}

We have presented an in-depth $^{75}$As NMR study of the critical
behavior of the SDW transition in the lightly Co doped regime of
Ba(Fe$_{1-x}$Co$_x$)$_2$As$_2$, with the primary focus on $x=0.02$.
We identified the NMR signatures of the tetragonal to orthorhombic
structural phase transition preceding the SDW transition.  Our
Knight shift, NMR line width, and $1/T_{1}$ data suggest that the strong
short range SDW order with three dimensional nature sets in
once the FeAs planes lower the symmetry from tetragonal to
orthorhombic.   In the orthorhombic phase below $T_{s}$,
simplistic fits of the antiferromagnetic contribution to $1/T_{1}T$
based on a Curie-Weiss law using two free parameters (Fig.~7) or 2D
SCR theory with four free parameters \cite{NakaiPRB} fail to capture
the critical behavior.  Precisely at $T_s$, critical slowing down of
spin fluctuations sets in, and the critical exponent for the
divergence of the dynamical structure factor $S({\bf q},
{\omega_{n}})$ is $\delta \sim  0.33$, as generally expected for
insulating 3D Heisenberg antiferromagnets.  Our fitting range is
rather limited and it is difficult to draw a definitive conclusion,
but this value is inconsistent with $\delta = 0.5$ expected for the
3D SCR theory for itinerant antiferromagnets \cite{Moriya}.

\section{ACKNOWLEDGEMENT}

The work at Zhejiang was supported by National Basic Research Program of China (No.2014CB921203,2011CBA00103), NSF of China (No. 11274268). The work at McMaster was supported by NSERC and CIFAR. The work at Oak Ridge National Laboratory was supported by the Department of Energy, Basic Energy Sciences, Materials Sciences and Engineering Division. The work at Beijing and Nanjing was supported by NSFC, the Ministry of Science and Technology of China, and the Chinese Academy of Sciences. \\

\end{document}